\def\ra{\rangle}
\def\la{\langle}
\def\be{\begin{equation}}
\def\ee{\end{equation}}
\def\ba{\begin{array}}
\def\ea{\end{array}}

\documentclass[aps,pre,preprint,showpacs,amsmath,amssymb,amsfonts,preprintnumbers]
{revtex4}
\usepackage{epsfig}

\begin{document}

\baselineskip=18pt \setcounter{page}{1} \centerline{\large\bf Upper
Bound of Fully Entangled Fraction } \vspace{4ex}
\begin{center}
Ming Li$^{1}$, Shao-Ming Fei$^{1,2}$ and  Zhi-Xi Wang$^{1}$

\vspace{2ex}

\begin{minipage}{5in}

\small $~^{1}$ {\small Department of Mathematics, Capital Normal
University, Beijing 100037}

{\small $~^{2}$ Institut f\"ur Angewandte Mathematik, Universit\"at
Bonn, D-53115}

\end{minipage}
\end{center}

\begin{center}
\begin{minipage}{5in}
\vspace{1ex} \centerline{\large Abstract} \vspace{1ex} We study the
fully entangled fraction of quantum states. An upper bound is
obtained for arbitrary dimensional bipartite systems. This bound is
shown to be exact for the case of two-qubit systems. An inequality
related the fully entangled fraction of two qubits in a three-qubit
mixed state has been also presented.

\smallskip
PACS numbers: 03.67.-a, 02.20.Hj, 03.65.-w\vfill
\smallskip
\end{minipage}\end{center}
\bigskip

The fully entangled fraction ($FEF$) is tightly related to
many quantum information processing such as
dense coding \cite{dc}, teleportation \cite{tel}, entanglement swapping
\cite{es}, and quantum cryptography (Bell inequalities) \cite{crypto}.
As the optimal fidelity of teleportation is given by $FEF$
\cite{horodecki}, experimentally measurement of $FEF$ can be also used to
determine the entanglement of the non-local source used in teleportation.
Thus an analytic formula for $FEF$ is of great importance.
In \cite{grondalski} an elegant formula for two-qubit
system is derived analytically by using the method of Lagrange
multipliers. Concerning
the estimation of entanglement of formation and concurrence,
exact results have been obtained not only for two-qubit case,
but also for some high dimensional states, isotropic and Werner states.
And analytical lower bounds have been obtained for general cases \cite{h4}.
While the analytical computation of $FEF$ remains formidable and less result has been known
for high dimensional quantum states.

In this paper, we study the fully entangled fraction of arbitrary
dimensional quantum bipartite states: the upper bound of $FEF$, its
relations to the filtering operations in the generalized
distillation protocol of entanglement, the relations between $FEF$
of two qubits in a three-qubit mixed state and the related
concurrence.

Let ${\mathcal {H}}$ be a $d$-dimensional complex vector space with computational
basis $|i\ra$, $i=1,...,d$.
The fully entangled fraction of a density matrix
$\rho\in{\mathcal {H}}\otimes{\mathcal {H}}$ is defined by
\begin{eqnarray}\label{def}
{\mathcal {F}}(\rho)=\max_{U}\la\psi_{+}|(I\otimes U^{\dag})\rho
(I\otimes U)|\psi_{+}\ra
\end{eqnarray}
under all unitary transformations $U$, where
$|\psi_{+}\ra=\frac{1}{\sqrt{d}}\sum\limits_{i=1}^{d}|ii\ra$
is the maximally entangled states and $I$ is the corresponding
identity matrix.

Let $\lambda_{i}$, $i=1,...,d^2-1$, be the generators of the $SU(d)$
algebra with $Tr \{\lambda_{i}\lambda_{j}\}=2\delta_{ij}$. A
bipartite state $\rho\in{\mathcal {H}}\otimes{\mathcal {H}}$ can be
expressed as
\begin{eqnarray}\label{rho}
\rho=\frac{1}{d^{2}}I\otimes
I+\frac{1}{d}\sum\limits_{i=1}^{d^{2}-1}r_{i}(\rho)\lambda_{i}\otimes
I+\frac{1}{d}\sum\limits_{j=1}^{d^{2}-1}s_{j}(\rho)I\otimes
\lambda_{j}+\sum\limits_{i,j=1}^{d^{2}-1}m_{ij}(\rho)\lambda_{i}\otimes
\lambda_{j},
\end{eqnarray}
where $r_{i}(\rho)=\frac{1}{2}Tr\{\rho\lambda_{i}(1)\otimes I\}$,
$s_{j}(\rho)=\frac{1}{2}Tr\{\rho I\otimes \lambda_{j}(2)\}$ and
$m_{ij}(\rho)=\frac{1}{4}Tr\{\rho \lambda_{i}(1)\otimes
\lambda_{j}(2)\}$. Let $M(\rho)$ denote the correlation matrix with
entries $m_{ij}(\rho)$.

{\bf{Theorem 1:}} For any $\rho\in{\mathcal
{H}}\otimes{\mathcal {H}}$, the fully entangled
fraction ${\mathcal {F}}(\rho)$ satisfies
\begin{eqnarray}\label{inequ}
{\mathcal {F}}(\rho)\leq
\frac{1}{d^{2}}+4||M^{T}(\rho)M(P_{+})||_{KF},
\end{eqnarray}
where $M^{T}$ stands for the transpose of $M$ and
$||M||_{KF}=Tr\sqrt{MM^{\dag}}$ is the Ky Fan norm of $M$.

{\bf{Proof:}} First, we note that
\begin{eqnarray*}
P_{+}=\frac{1}{d^{2}}I\otimes
I+\sum\limits_{i,j=1}^{d^{2}-1}m_{ij}(P_{+})
\lambda_{i}\otimes \lambda_{j},
\end{eqnarray*}
where $m_{ij}(P_{+})=\frac{1}{4}Tr\{P_{+}
\lambda_{i}\otimes \lambda_{j}\}$.

By definition ($\ref{def}$), one obtains
\begin{eqnarray*}\label{pro}
{\mathcal {F}}(\rho)&=&\max_{U}\la\psi_{+}|(I\otimes U^{\dag})\rho
(I\otimes U)|\psi_{+}\ra \\
&=&\max_{U}Tr\{\rho (I\otimes U) P_{+} (I\otimes U^{\dag})\}\\
&=&\max_{U}[\frac{1}{d^{2}}Tr\{\rho\}+\sum\limits_{i,j=1}^{d^{2}-1}m_{ij}(P_{+})
Tr\{\rho\lambda_{i}\otimes U\lambda_{j}U^{\dag}\}].
\end{eqnarray*}

Since $U\lambda_{i}U^{\dag}$ is a traceless Hermitian operator, it
can be expanded according to the $SU(d)$ generators,
\begin{eqnarray}\label{rr}
U\lambda_{i}U^{\dag}=\sum\limits_{j=1}^{d^{2}-1}\frac{1}{2}Tr\{U\lambda_{i}U^{\dag}\lambda_{j}\}\lambda_{j}\equiv
\sum\limits_{j=1}^{d^{2}-1}O_{ij}\lambda_{j}.
\end{eqnarray}
Entries $O_{ij}$ defines a real $(d^{2}-1)\times(d^{2}-1)$ matrix $O$. From the
completeness relation of $SU(d)$ generators
\begin{eqnarray}
\sum\limits_{j=1}^{d^{2}-1}(\lambda_{j})_{ki}(\lambda_{j})_{mn}=2\delta_{im}\delta_{kn}
-\frac{2}{d}\delta_{ki}\delta_{mn},
\end{eqnarray}
one can show that $O$ is an orthonormal matrix. Using (\ref{rr}) we have
\begin{eqnarray*}\label{propp}
{\mathcal {F}}(\rho)&\leq&
\frac{1}{d^{2}}+\max_{O}\sum\limits_{i,j,k}m_{ij}(P_{+})O_{jk}Tr\{\rho\lambda_{i}\otimes
\lambda_{k}\}\\
&=&\frac{1}{d^{2}}+4\max_{O}\sum\limits_{i,j,k}m_{ij}(P_{+})O_{jk}m_{ik}(\rho)\\
&=&\frac{1}{d^{2}}+4\max_{O}Tr\{M(\rho)^{T}M(P_{+})O\}\\
&=&\frac{1}{d^{2}}+4||M(\rho)^{T}M(P_{+})||_{KF}.
\end{eqnarray*}
\hfill$\Box$

For the case $d=2$, we can get an exact result from (\ref{inequ}):

{\bf{Corollary:}} For two qubits system, we have
\be\label{corollary}
{\mathcal{F}}(\rho)=\frac{1}{4}+4||M(\rho)^{T}M(P_{+})||_{KF},
\ee
i.e. the upper bound derived in Theorem 1 is exactly the $FEF$.

{\bf{Proof:}} We have shown in ($\ref{rr}$) that given an arbitrary
unitary $U$, one can always obtain an orthonormal matrix $O$. Now we
show that in two-qubit case, for any $3\times3$ orthonormal matrix
$O$ there always exits $2\times2$ unitary matrix $U$ such that
(\ref{rr}) holds.

For any vector ${\bf{t}}=\{t_{1},t_{2},t_{3}\}$ with unit norm,
define an operator $X\equiv\sum\limits_{i=1}^{3}t_{i}\sigma_{i},$
where $\sigma_{i}$s are Pauli matrices. Given an orthonormal matrix
$O$ one obtains a new operator
$X^{'}\equiv\sum\limits_{i=1}^{3}t_{i}^{'}\sigma_{i}=\sum\limits_{i,j=1}^{3}O_{ij}t_{j}\sigma_{i}$.

$X$ and $X^{'}$ are both hermitian traceless matrices. Their eigenvalues
are given by the norms of the vectors ${\bf{t}}$
and ${\bf{t^\prime}}=\{t_{1}^\prime,t_{2}^\prime,t_{3}^\prime\}$ respectively.
As the norms are invariant under orthonormal transformations $O$,
they have the same eigenvalues: $\pm\sqrt{t_{1}^{2}+t_{2}^{2}+t_{3}^{2}}$.
Thus there must be a unitary matrix $U$ such that $X^{'}=UXU^{\dag}$.
Hence the inequality in the proof of Theorem 1 becomes an equality. The upper bound
($\ref{inequ}$) then becomes exact at this situation,
which is in accord with the result in \cite{grondalski}.
$\hfill\Box$

{\it Remark} The upper bound of $FEF$ (\ref{inequ}) and the $FEF$
(\ref{corollary}) for a state $\rho$ depend on the correlation
matrices $M(\rho)$ and $M(P_+)$. They can be calculated directly
according to a given set of $SU(d)$ generators $\lambda_{i}$,
$i=1,...,d^2-1$. Nevertheless the $FEF$
and its upper bound do not depend on the choice of the $SU(d)$ generators.

The upper bound can give rise to not only an estimation of the fidelity
in quantum information processing such as teleportation, but also an
interesting application in entanglement distillation of quantum states.
In \cite{gdp,rc}, a separability criterion called reduction criterion has been
proposed. It says that if a bipartite quantum state $\rho$ is
separable, then $(\rho_{1}\otimes I) -\rho \geq 0$ and $(I\otimes
\rho_{2})-\rho \geq 0$, where $\rho_{1}=Tr_2(\rho)$ (resp. $\rho_{2}=Tr_1(\rho)$) is
the reduced density matrix obtained by tracing over the second (resp. first)
subsystem. Here a matrix $X\geq 0$ means that all the eigenvalues of
$X$ are greater than or equal to $0$. In \cite{gdp}
a generalized distillation protocol has been presented. It is shown that a
quantum state $\rho$ violating the reduction criterion can always be
distilled. For such states if their single fraction of entanglement
$F(\rho)=\la\psi_{+}|\rho|\psi_{+}\ra$ is greater than
$\frac{1}{d}$, then one can distill these states directly
by using the generalized distillation protocol. However
if even the $FEF$ (the largest value of single fraction of entanglement under local
unitary transformations) is less than or equal to $\frac{1}{d}$, then a
proper filtering operation has to be used at first to transform
$\rho$ to another state $\rho^{'}$ so that $F(\rho^{'})>\frac{1}{d}$.
For $d=2$, one can compute $FEF$ analytically according to the
corollary. For $d \geq 3$ our upper bound ($\ref{inequ}$) can supply
a necessary condition in the distillation:

{\bf{Theorem 2:}} For an entangled state $\rho\in
{\mathcal{H}}\otimes{\mathcal {H}}$ violating the reduction criterion,
if the upper bound ($\ref{inequ}$) is less than or equal to
$\frac{1}{d}$, then the filtering operation has to be applied before
using the generalized distillation protocol.

As an example we consider a $3\times 3$ state
\begin{eqnarray}
\rho=\frac{8}{9}\sigma+\frac{1}{9}|\psi_{+}\ra\la\psi_{+}|,
\end{eqnarray}
where $\sigma=(x|0\ra\la 0|+(1-x)|1\ra\la 1|)\otimes(x|0\ra\la
0|+(1-x)|1\ra\la 1|)$. It is direct to verify that $\rho$ violates
the reduction criterion for $0\leq x\leq 1$, as $(\rho_{1}\otimes
I)-\rho$ has a negative eigenvalue $-\frac{2}{27}$. Therefore the
state is distillable. From Fig. 1 we see that for $0\leq x<0.0722$ and
$0.9278<x\leq 1$, the fidelity is already greater than
$\frac{1}{3}$, thus the generalized distillation protocol can be
applied without the filtering operation. However for $0.1188\leq x\leq
0.8811$, even the upper bound of the fully entangled fraction is
less than or equal to $\frac{1}{3}$, hence the filtering operation
has to be applied first, before using the generalized distillation protocol.

\begin{figure}[tbp]
\begin{center}
\resizebox{10cm}{!}{\includegraphics{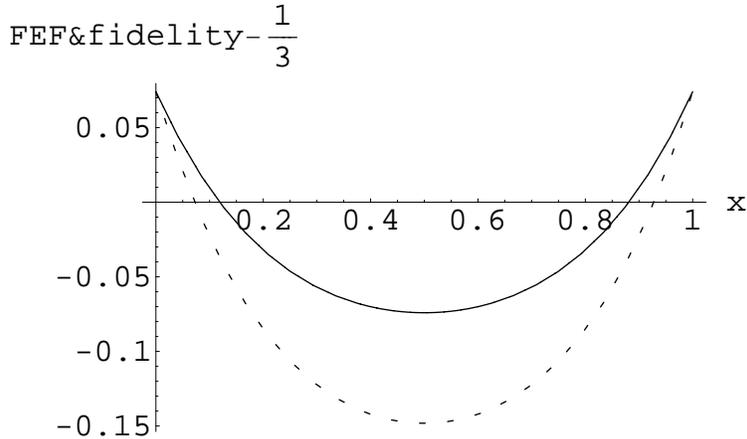}}
\end{center}
\caption{Upper bound of ${\mathcal {F}}(\rho)-\frac{1}{3}$ from
($\ref{inequ}$) (solid line) and fidelity $F(\rho)-\frac{1}{3}$
(dashed line). \label{fig1}}
\end{figure}

The upper bound of $FEF$ has also interesting relations to the entanglement
measure concurrence. Let us consider tripartite case. Let $\rho_{ABC}$ be a state of
three-qubit systems denoted by $A$, $B$ and $C$. We study
the upper bound of the $FEF$, ${\cal F}(\rho_{AB})$,
between qubits $A$ and $B$, and its relations to the concurrence
under bipartite partition $AB$ and $C$.
For convenience we normalize ${\cal F}(\rho_{AB})$ to be
\begin{eqnarray}\label{frho}
{\mathcal {F}}_{N}(\rho_{AB})=\max\{2{\mathcal {F}}(\rho_{AB})-1,0\}.
\end{eqnarray}
For a bipartite pure state $|\psi\ra\in{\mathcal {H}}\otimes{\mathcal {H}}$, the concurrence
\cite{concurrence} is defined by
$C(|\psi\ra)=\sqrt{2(1-Tr\{\rho_{1}^{2}\})}$.
The concurrence is extended to mixed states $\rho$ by the convex roof,
$C(\rho)\equiv\min\limits_{\{p_{i},|\psi_{i}\ra\}}\sum\limits_{i}p_{i}C(|\psi_{i}\ra)$
for all possible ensemble realizations
$\rho=\sum\limits_{i}p_{i}|\psi_{i}\ra\la\psi_{i}|$, $p_{i}\geq0$, $\sum\limits_{i}p_{i}=1$.
Let $C({\rho_{AB|C}})$ denote the concurrence between subsystems $AB$ and $C$.

\bigskip
{\bf{Theorem 3:}} For any triqubit state $\rho_{ABC}$, ${\mathcal
{F}}_{N}(\rho_{AB})$ satisfies \be\label{inequformix} {\mathcal
{F}}_{N}(\rho_{AB})\leq\sqrt{1-C^{2}({\rho_{AB|C}})}. \ee

{\bf{Proof:}} We first consider the case that $\rho_{ABC}$ is pure,
$\rho_{ABC}=|\psi\ra_{ABC}\la\psi|$. By using the Schmidt
decomposition between qubits $A, B$ and $C$,
$|\psi\ra_{ABC}$ can be written as:
\begin{eqnarray}\label{schmidt}
|\psi\ra_{AB|C}=\sum\limits_{i=1}^{2}\eta_{i}|i_{AB}\ra|i_{C}\ra,
\eta_{1}^{2}+\eta_{2}^{2}=1,~~~\eta_{1}\geq \eta_{2}
\end{eqnarray}
for some othonormalized bases $|i_{AB}\ra$, $|i_{C}\ra$ of subsystems $AB$, $C$ respectively.
The reduced density matrix $\rho_{AB}$ has the form
\begin{eqnarray*}
\rho_{AB}&=&Tr_{C}\{\rho_{ABC}\}=\sum\limits_{i=1}^{2}\eta_{i}^{2}
|i_{AB}\ra\la i_{AB}|=U^{T}\Lambda U^{*},
\end{eqnarray*}
where $\Lambda$ is a $4\times 4$ diagonal matrix with diagonal
elements $\{\eta_{1}^{2}, \eta_{2}^{2}, 0, 0\}$, $U$ is a unitary matrix and
$U^{*}$ denotes the conjugation of $U$.

The $FEF$ of the two-qubit state $\rho_{AB}$ can be calculated by
using formula (\ref{corollary}) or the one in \cite{grondalski}. Let
$$M=\frac{1}{\sqrt{2}}\left(%
    \begin{array}{cccc}
      1 & 0 & 0 & i \\
      0 & i & -1 & 0 \\
      0 & i & 1 & 0 \\
      1 & 0 & 0 & -i \\
    \end{array}%
    \right)
    $$
be the $4\times 4$ matrix constituted by the four Bell bases. The
$FEF$ of $\rho_{AB}$ can be written as
\be\ba{rcl}\label{pppp}
{\mathcal {F}}(\rho_{AB})&=&\eta_{max}(Re\{M^{\dag}\rho_{AB} M\})
=\frac{1}{2}\eta_{max}(M^{\dag}\rho_{AB} M+M^{T}\rho_{AB}^{*} M^{*}) \\
&\leq&\frac{1}{2}[\eta_{max}(M^{\dag}U^{T}\Lambda U^{*}M)+\eta_{max}(M^{T}U^{\dag}\Lambda UM^{*})]
=\eta_{1}^{2}
\ea
\ee
where $\eta_{max}(X)$ stands for the maximal eigenvalues of the matrix $X$.

For pure state ($\ref{schmidt}$)
in bipartite partition $AB$ and $C$, we have
\begin{eqnarray}\label{con}
C(|\psi\ra_{AB|C})=\sqrt{2(1-Tr\{\rho_{AB}^{2}\})}=2\eta_{1}\eta_{2}.
\end{eqnarray}
From (\ref{frho}), (\ref{pppp}) and (\ref{con}) we get
\begin{eqnarray}\label{inequality}
{\mathcal {F}}_{N}(\rho_{AB})\leq\sqrt{1-C^{2}(|\psi\ra_{AB|C})}.
\end{eqnarray}

We now prove that the above inequality ($\ref{inequality}$) also
holds for mixed state $\rho_{ABC}$. Let
$\rho_{ABC}=\sum\limits_{i}p_{i}|\psi_{i}\ra_{ABC}\la\psi_{i}|$ be
the optimal decomposition of $\rho_{ABC}$ such that
$C(\rho_{AB|C})=\sum\limits_{i}p_{i}C(|\psi_{i}\ra)_{AB|C}$. We have
\begin{eqnarray*}
{\mathcal {F}}_{N}(\rho_{AB})
&\leq&\sum\limits_{i}p_{i}{\mathcal {F}}_{N}({\rho^{i}_{AB}})
\leq\sum\limits_{i}p_{i}\sqrt{1-C^{2}({\rho^{i}_{AB|C}})}\\
&\leq&\sqrt{1-\sum\limits_{i}p_{i}C^{2}({\rho^{i}_{AB|C}})}
\leq\sqrt{1-C^{2}({\rho_{AB|C}})},
\end{eqnarray*}
where $\rho^{i}_{AB|C}=|\psi_{i}\ra_{ABC}\la\psi_{i}|$ and
$\rho^{i}_{AB}=Tr_{C}\{\rho^{i}_{AB|C}\}$.
$\hfill\Box$

\begin{figure}[tbp]
\begin{center}
\resizebox{10cm}{!}{\includegraphics{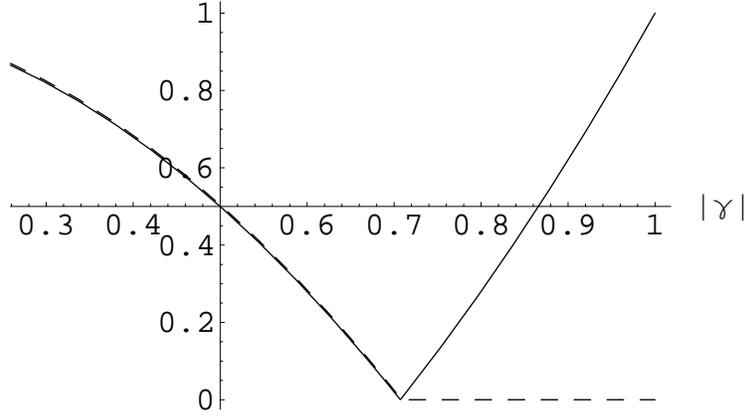}}
\end{center}
\caption{${\mathcal {F}}_{N}(\rho_{AB}^{W^{'}})$ (dashed line) and
Upper bound $\sqrt{1-C^{2}(|W^{'}\ra_{AB|C})}$ (solid line) of state
$|W^{'}\ra_{AB|C}$ at $|\alpha|=|\beta|$. \label{fig2}}
\end{figure}

From Theorem 2 we see that the $FEF$ of quibts $A$ and $B$ are
bounded by the concurrence between qubits $A$, $B$ and qubit $C$.
The upper bound of $FEF$ for $\rho_{AB}$ decreases when the
entanglement between qubits $A, B$ and $C$  increases.
As an example, we consider the generalized W state defined by
$|W^{'}\ra=\alpha|100\ra+\beta|010\ra+\gamma|001\ra$,
$|\alpha|^{2}+|\beta|^{2}+|\gamma|^{2}=1$.
The reduced density matrix is given by
$$\rho_{AB}^{W^{'}}=\left(%
    \begin{array}{cccc}
      |\gamma|^{2} & 0 & 0 & 0 \\
      0 & |\beta|^{2} & \alpha^{*}\beta & 0 \\
      0 & \alpha\beta^{*} & |\alpha|^{2} & 0 \\
      0 & 0 & 0 & 0 \\
    \end{array}%
    \right).
    $$
The $FEF$ of $\rho_{AB}^{W^{'}}$ is given by
$$
{\mathcal
{F}}_{N}(\rho_{AB}^{W^{'}})=-\frac{1}{2}+2|\alpha||\beta|+\frac{1}{2}||\alpha|^{2}+|\beta|^{2}-|\gamma|^{2}|.
$$
While the concurrence of $|W^{'}\ra$ has the from
$C_{AB|C}(|W^{'}\ra)=2|\gamma|\sqrt{|\alpha|^{2}+|\beta|^{2}}$.
We see that (\ref{inequformix}) always holds. In particular for
$|\alpha|=|\beta|$ and $|\gamma|\leq \frac{\sqrt{2}}{2}$, the
inequality (\ref{inequformix}) is saturated (see Fig. 2).

\medskip

We have studied the fully entangled fraction of arbitrary
dimensional quantum bipartite states. We obtained an analytic upper
bound of $FEF$, which is exact the $FEF$ for two-qubit systems. This
upper bound of $FEF$ gives a necessary condition for which the
filtering step has to be performed in the generalized distillation
protocol of entanglement. An inequality related the fully entangled
fraction of two qubits in a three-qubit mixed state has been also
presented. As the fully entangled fraction is directly related to
dense coding, teleportation, entanglement swapping and quantum
cryptography, the results could shed new lights on the study of
relevant quantum information processing both theoretically and
experimentally.

\bigskip
\noindent{\bf Acknowledgments}\, This work is supported by NSFC
under grant 10675086, NKBRSFC under grant 2004CB318000.

\end{document}